\documentclass{article}
\PassOptionsToPackage{numbers, compress}{natbib}

\usepackage[preprint]{neurips_2023}

\usepackage[utf8]{inputenc} 
\usepackage[T1]{fontenc}    
\PassOptionsToPackage{hyphens}{url}
\usepackage{url}            
\usepackage{hyperref}       
\usepackage{booktabs}       
\usepackage{amsfonts}       
\usepackage{nicefrac}       
\usepackage{microtype}      
\usepackage{xcolor}         
\usepackage[capitalise]{cleveref}

\usepackage[hyphenbreaks]{breakurl}

\title{An Alternative to Regulation: The Case for Public AI}
\author{
  Nicholas Vincent \\
  Simon Fraser University \\
  Computing Science \\
  \texttt{nvincent@sfu.ca} \\
  \And
  David Bau \\
  Northeastern University \\
  Khoury College of Computer Sciences \\
  \texttt{davidbau@northeastern.edu} \\
  \And
  Sarah Schwettmann \\
  Massachusetts Institute of Technology \\
  Computer Science and Artificial Intelligence Lab \\
  \texttt{schwett@mit.edu} \\
  \And
  Joshua Tan \\
  University of Oxford and Metagov \\
  \texttt{joshua.tan@magd.ox.ac.uk} \\
}

\begin{document}

\maketitle

\begin{abstract}
Can governments build AI? In this paper, we describe an ongoing effort to develop ``public AI''---publicly accessible AI models funded, provisioned, and governed by governments or other public bodies. Public AI presents both an alternative and a complement to standard regulatory approaches to AI, but it also suggests new technical and policy challenges. We present a roadmap for how the ML research community can help shape this initiative and support its implementation, and how public AI can complement other responsible AI initiatives.
\end{abstract}

\section{Introduction}
In light of continued progress in AI, there is growing concern about the status quo in which powerful consumer-facing AI systems are operated almost exclusively by a small number of private firms, who primarily build and deploy AI systems using privately owned computing infrastructure (see e.g. discussion such as \cite{vincentAIEnteringEra2023,widderOpenBusinessBig2023,sadowski_everyone_2021}).
In response, a number of voices have called for public bodies to directly support the development and deployment of public interest AI models \cite{jenningsOpinionThereOnly2023,schneierHowArtificialIntelligence2023,sandersBuildAIPeople2023}. In the UK and Sweden, there are early efforts to organize country-level ChatGPT-like systems \cite{belfieldGreatBritishCloud2023, ukparliamentGovernanceArtificialIntelligence2023,AISwedenAdvancing2023}. In the US, there are efforts to build a National AI Research Resource (NAIRR) \cite{lynchNewReportDetails2023}, and a recent executive order signaled support for both government use of and investment in AI that serves the public good \cite{thewhitehouseFACTSHEETPresident2023}.

These efforts are early steps towards \emph{public AI}---publicly accessible AI models funded, provisioned, and governed by governments or other public bodies. 
Public AI might take the form of a ``public option'' service for large language models (like a public option for banking), a national AI agency (like a national health service), or a set of library-like organizations that offer self-serve access to and support for a variety of models.
In any of its instantiations, public AI offers a realistic institutional alternative to an ecosystem where AI models, especially foundation models \cite{bommasani2021opportunities}, are primarily maintained by private corporate actors (with or without regulation). As an approach for provisioning access to AI, public AI also offers an alternative to open-source AI and decentralized AI.

Public AI projects share a goal of creating AI systems that are built, governed, and operated in accordance with shared values and for the public’s benefit. They also provide a mechanism to ensure public funding for AI research results into public benefit rather than private capture. Ethically, developing public AI reflects the shared, open, and public nature of the internet and of the cultural data upon which AI foundation models are built \cite{huangGenerativeAIDigital2023,liDimensionsDataLabor2023}. For a longer rationale, we defer to \cite{publicai}.

In this paper we describe the public AI concept, comparing it to other regulatory approaches. We then propose ways for the research community to contribute to and make use of public AI, organized in terms of open challenges and new opportunities for research. In other words, we explain how public AI motivates exciting new research questions, and why public AI can directly benefit researchers in academic and non-academic settings.

\section{Background}
There is a long-running tradition of public bodies funding and otherwise supporting emerging technologies. In this sense, public AI is an extension of the logic that has engaged governments in high-stakes technological development \cite{dolfsmaGovernmentPolicyTechnological2013}.

A variety of ongoing initiatives that range from nascent ideas to fleshed-out policy proposals aim to build something close to the definition of \emph{public} we use here---AI technologies that are publicly accessible, publicly funded and/or provisioned, and governed by public bodies accountable to the public. This definition permits a wide range of approaches that span different processes for provisioning AI outputs, auditing and updating models, and funding and handling operational infrastructure challenges in collaboration with private and nonprofit entities. ``AI models'' can refer to a broad class of data-dependent systems, but our discussion specifically centers foundation models \cite{bommasani2021opportunities}. However, the positive outcomes from investing in public AI (especially with respect to building state capacity) will be synergistic with public use of simpler ``AI'' systems.

Support for public AI has appeared in a range of venues. In 2019, \citeauthor{ganskyOpinionArtificialIntelligence2019} called for a public research consortium in a New York Times op-ed \cite{ganskyOpinionArtificialIntelligence2019}. More recently, pieces in Politico \cite{jenningsOpinionThereOnly2023}, Slate \cite{schneierHowArtificialIntelligence2023}, and more \cite{sandersBuildAIPeople2023} have argued for public AI. In the US, the proposed National Artificial Intelligence Research Resource Task Force (NAIRR) serves as an example of shared computing and data infrastructure \cite{lynchNewReportDetails2023}. Furthermore, an executive order on AI issued in November 2023 and the release of the ``ai.gov'' website signaled further support for public AI-type initiatives, including efforts by the government to hire AI talent, use AI, and support democratic inputs in AI policy. In the UK, Parliament has expressed serious interest in supporting AI-related initiatives \cite{ukparliamentGovernanceArtificialIntelligence2023}.

In the coming months we expect to see discussion of public AI advance in both public forums (e.g., high profile op-eds) and in government bodies (e.g., the U.S. and U.K. lawmaking bodies). We also expect these discussions identify weaknesses of public AI, including both contexts where public systems might struggle to compete with private alternatives and contexts in which public bodies may fail to meet standards of transparency and participation. We hope that the ML community can help lead these discussions, for example by organizing concrete evidence on potential harms or by working with governments to build and govern initial pilots.

\subsection{Comparison to other regulatory approaches}
There are a variety of regulatory modalities that governments can already apply to AI: direct regulation that restricts the behavior of AI companies and developers, trade policy \cite{irion_ai_2021, tan_how_2023}, competition policy, tax incentives, and direct grants and contracts.
Within this space of regulatory interventions, public AI has a number of distinct advantages and drawbacks.

First and foremost, the act of building and provisioning AI provides institutions with indispensable practical experience---experience that does not arrive from hands-off regulation. In this view, public AI is a direct complement to other regulatory actions insofar as it builds the expertise and capacity to regulate within public bodies. Even compared to a ``heavy'' regulatory regime, where a given jurisdiction regulates AI companies akin to (privately-owned) public utilities, there is no replacement for actually building AI.

Second, public AI centers a particular expertise of governments: vendor selection. Governments contract significant amounts of work to other bodies, whether private entities, nonprofits, or academic institutions (cf. research grants). While not all forms of public AI emphasize contracting (e.g. opting for a national AI agency), governments are uniquely positioned to bring together organizations and stakeholders from a wide range of industries within a shared enterprise, e.g. the COVID-19 vaccines or the human genome project. Public AI thus represents one channel for deploying resources arranged for AI development.

Third, deploying a model or service comes with its own costs and risks. Perhaps the biggest risk: public AI has to be \emph{good}, or at least competitive with private options, or risk irrelevance---though it may still serve to develop government expertise, provide a platform for science (see \cref{sec:transparency}), or help to develop standards (see \cref{sec:standards}). 
Note that the riskiness of adoption depends on the availability of alternatives (private, open-source, and other) within each particular jurisdiction as well as the particular form of public AI, e.g. some versions of public AI may use privately-developed or open-source models. 
By offering a service, public bodies may open themselves up to liability and other costs and claims (though how liability applies to some generative AI use cases remains to be decided \cite{walshLegalIssuesPresented2023}) . While governments are no strangers to service provision, some may face the cost of learning to grapple with these concerns  in a new domain for the first time. Direct regulation does not directly incur these costs.

Lastly, public AI can be more participatory than direct regulation (see e.g. \citeauthor{delgadoParticipatoryTurnAI2023} \cite{delgadoParticipatoryTurnAI2023} for a discussion of participatory AI). The drafting of industry regulations is typically a technical, elite activity. Public AI, by virtue of its accessibility, offers a direct line by which a citizenry can engage with and see the effects of government action. Many parties have called for ``democratic input'' to AI, but most proposals assume such input to the development of otherwise undemocratic private models \cite{zarembaDemocraticInputsAI2023}. Public AI has an enforceable form of democratic input built in, though the nature of that input depends on the governance model of the jurisdiction and on practical decisions around implementation---e.g. whether public hearings, learning from citizen interactions, or just a plain vote.

\section{Open challenges for the research community}
Implementing public AI poses unique technical and policy challenges; it also offers a new perspective on existing research questions within AI safety, interpretable ML, and technology governance more broadly. The research community has a critical role to play in posing and solving these challenges.

\subsection{Data practices}
There are many open questions that will influence how public AI institutions obtain and use data. For example: how will the privacy of information in training data be managed? Must personally identifiable data be removed, as faces were blurred in ImageNet~\cite{yang2022faceobfuscation}? What limits are there on objectionable content in training data?
Stable Diffusion 2 removed nudity from its training data~\cite{rombach2022sd20}. Must the intellectual property used to create public AI also be public? Efforts such as SpawningAI~\cite{spawningai} are creating systems to allow copyright holders to remove their content from training data. 

These questions may be answered based on research about data valuation and people's preferences regarding data use. Should public AI be subject to the same rules as private AI, or might it have broader rights, for instance better arguments around fair use claims that hinge on serving the public interest?

Technical methods also need to be developed to enable attribution and editing of data in response to policy choices and regulation. For example, when an improper piece of training data is identified, it may be impractical for public AI systems to be fully retrained with that data item removed, potentially at massive expense. Public AI will be a natural application for efficient methods for handling responsible data curation~\citep{meng2022locating,meng2022mass,hartvigsen2022aging,gandikota2023unified}.

Finally, there is the question of how these decisions should be made. In private organizations, there is no guarantee of consistency, transparency, or fairness; public institutions have stronger norms around this but must engage in decision-making appropriate for the fast-changing environment.  

\subsection{Models for public-private partnership}

One of the ongoing challenges that will face public AI will be to ensure that public models employ state-of-the-art methods, even as the state-of-the-art advances quickly. Private companies are making tremendous investments in AI technology, so a public AI will be most useful if it is able to benefit from private-sector innovations.

Yet if private companies participate in the creation of public AI, we will need to ask: to what extent can they contribute while protecting their private intellectual property?  There is a possible distinction between the methods and data used to train an AI and the design, parameters, and knowledge contained within the AI itself.  And participants in a public AI project may choose to contribute some methods while holding others back. Large-scale open-source projects provide one possible inspiration, where many private companies make individual decisions about which intellectual property to contribute and which to hold back as proprietary. A framework for similar collaborative engineering in public machine learning will be a key ingredient for public AI.

\subsection{Engineering challenges}

The same model with the same weights may be public, private, or open-source. But public AI is not just a form of private, corporate AI where we swap all the `Anthropic' labels in the Terms of Service with `Department of Commerce' or `State of Pennsylvania'---nor is it just open-source AI with a public body as the controlling entity of the repository rather than a foundation or private company. For example, in the interest of accessibility, public operators may put more emphasis on offering \emph{some} level of service rather than a consistent product offering---but how do we offer a lower-cost experience that degrades gracefully while preserving safety? Or, suppose that public bodies want to publish summary statistics in the interest of accountability, but not all model weights---what statistics should they publish, and how can we help the public interpret this technical information?

For now, many of the engineering challenges are still to be discovered. To understand the scope of these challenges, consider a comparison: open-source AI in its current incarnation requires a wide range of new technologies and services to be practically workable compared to private AI---licenses, model hubs, infrastructure for running and deploying models, community management, and more.
The engineering challenges faced by a community of open-source AI contributors are not the same as those faced by a group of engineers operating within a private company because of disparities in engineering resources (data, compute, know-how, management, etc.). Nor will they be the same engineering challenges faced by the builders of public AI.

\subsection{Preventing misuse}
As large AI models become more capable, the risk of misuse will also grow. Highly-capable AI models could be used to create persuasive misinformation~\cite{torzdl2019faceswap,mirsky2021creation,zhou2023synthetic,gao2023comparing,gravel2023learning}, create dangerous substances~\cite{soice2023can,bran2023chemcrow,guo2023indeed}, or to find vulnerabilities in other systems~\cite{deng2023pentestgpt,gupta2023chatgpt}. Public AI must have a mechanism to limit these types of applications in a way that honors the principle of public accessibility.

There are two natural approaches to ensure that a system is used in ways that advance the public interest. One is to regulate and monitor \emph{usage} of the system; and the other is govern the \emph{creation} of the system. Both present policy and research challenges; public AI emphasises the latter. This is particularly salient given the ongoing challenges of monitoring private models operating on private devices.

\section{Opportunities for the research community}

\subsection{Research}
First and foremost, support for public AI can help allocate funding and other important kinds of capital to a variety of public-interest AI research projects. A successful national implementation of public AI would operate at a far greater scale---and afford many more opportunities for research---than even the largest public funding proposals being discussed right now.

Public AI would likely be developed in partnership with academic institutions and offer teaching and research opportunities, akin to the role of CERN for experimental physics. Public AI could especially complement ``talent surges'' for government hiring in AI \cite{thewhitehouseFACTSHEETPresident2023}. Furthermore, public AI offers an additional path to help research conducted in academic settings reach the front lines and directly contribute to public good. 

Finally, concrete steps to support public AI can help lend credence to public interest research. Even for researchers that don't directly receive resources as a result of public AI programs, by legitimizing an alternative future, it may be easier to publish and share work. Creating a sense of a shared mission -- with concrete progress -- can motivate public interest research and help those conducting it justify their work.

\subsection{Transparency} \label{sec:transparency}
Public AI will support transparent studies of mechanisms of very large models, beyond what might be achieved with a focus on open-source (but privately steered) development \cite{widderOpenBusinessBig2023}. Open code and open weights do not guarantee transparent access to model outputs, but publicly provisioned models can.

Currently, researchers with expertise in scientifically evaluating model behavior may not have access to the internals of proprietary models that may be crucial for developing understanding. Furthermore, there are major challenges in using traditional empirical methods (e.g. audits) to study private large models. In the wake of increased concerns around competition and safety, firms now reveal extremely minimal details about model design (does it use an ensemble? what layers of filtering and classification are applied to prompts? etc.). Carefully collected audit-style data may be misinterpreted. Public AI enables better study of core model capabilities, which is crucial for downstream studies of societal impact. Another possible advantage of public AI may be working with census bodies to develop truly representative ecologically valid evaluation procedures.

Finally, it is important to note that transparency for some models may be in direct tension with safety. For these cases, public auditing can help, but it may be the case that maximal transparency is not always desirable. Some kind of gradation of transparency could be achieved via the accountability component of public AI.

\subsection{Accelerating standardization}\label{sec:standards}

Public AI can accelerate standardization in AI---a major focus of the recent US executive order \cite{thewhitehouseFACTSHEETPresident2023}---by providing exemplar systems. Perhaps more importantly, insofar as public AI defines a floor for AI services and a ``base level'' of AI services in a given jurisdiction, it can prevent both common abuses of market power as well as the selective inattention experienced by Global South countries that arise when a single company controls a widely-used platform.

Successful public AI projects will also advance the development of a set of standard technological components for AI. In a world with standardized components (defined and maintained by publicly accountable standards bodies), both public institutions aiming to make scientific progress and private institutions aiming to build capable consumer-facing AI products will benefit.

Standardization can be decisive in increasing the stability of downstream product development. If an industry becomes heavily reliant on AI model outputs (e.g., suppose the medical text processing industry comes to rely on a specific AI model), they may be vulnerable to massive disruption if one or two private AI operators change their policies.

\section{Conclusion}
Here, we have described the public AI concept as an alternative and complement to regulation. We have focused highlighted both open challenges and reasons the research community might be excited about this approach. We refer readers to \url{https://publicai.network/} for more information about public AI-related initiatives.

\section*{Acknowledgements}
We would like to thank SJ Klein for helpful comments and feedback on earlier drafts of this article. We would also like to thank many participants in various public AI-related discussions for their contributions.

\setcitestyle{numbers}
\bibliographystyle{plainnat}
\bibliography{bib}

\end{document}